\documentclass[12pt]{article}
\usepackage{amsmath}
\usepackage{graphicx,psfrag,epsf}
\usepackage{enumerate}
\usepackage{bm}
\usepackage{bigints}
\usepackage{xcolor}
\usepackage{natbib}
\usepackage{url} 
\usepackage{setspace}

\newcommand{\blind}{0}
\usepackage{amsfonts}
\newcommand{\mb}[1]{\mathbf{0}}

\begin{document}

\def\spacingset#1{\renewcommand{\baselinestretch}%
{#1}\small\normalsize} \spacingset{1}


\if0\blind
{
  \title{\bf Assessing Bayes factor surfaces using interactive visualization and computer surrogate modeling}
  \author{Christopher T. Franck\\
    Department of Statistics, Virginia Tech\\
    and \\
    Robert B. Gramacy \\
    Department of Statistics, Virginia Tech}
  \maketitle
} \fi

\if1\blind
{
  \bigskip
  \bigskip
  \bigskip
  \begin{center}
    {\LARGE\bf Title}
\end{center}
  \medskip
} \fi

\bigskip
\begin{abstract}
Bayesian model selection provides a natural alternative to classical hypothesis testing based on $p$-values. While many papers mention that Bayesian model selection can be sensitive to prior specification on parameters, there are few practical strategies to assess and report this sensitivity. This article has two goals.  First, we aim to educate the broader statistical community about the extent of potential sensitivity through visualization of the Bayes factor surface.  The Bayes factor surface shows the value a Bayes factor takes as a function of user-specified hyperparameters. Second, we suggest surrogate modeling via Gaussian processes to visualize the Bayes factor surface in situations where computation is expensive. We provide three examples including an interactive {\sf R} {\tt shiny} application that explores a simple regression problem, a hierarchical linear model selection exercise, and finally surrogate modeling via Gaussian processes to a study of the influence of outliers in empirical finance. We suggest Bayes factor surfaces are valuable for scientific reporting since they (i) increase transparency by making instability in Bayes factors easy to visualize, (ii) generalize to simple and complicated examples, and (iii) provide a path for researchers to assess the impact of prior choice on modeling decisions in a wide variety of research areas.

\end{abstract}

\noindent%
\vfill

\newpage
\setstretch{1.45}
\section{Introduction}
\label{sec:intro}

In the current scientific landscape, multiple concerns surrounding the use of classical $p$-values for null hypothesis significance testing have been raised, see e.g. \cite{Ioannidis2005}, \cite{Boos2011}, \cite{Young2011}, \cite{Trafimow2015}, \cite{Wasserstein2016}, and the recent American Statistician Special Issue: {\em Statistical Inference in the 21st Century: A World Beyond $p<0.05$}. Despite these concerns, interest in hypothesis testing persists.  Bayesian model selection  enables direct probabalistic statements about the plausibility of competing models and is thus a natural alternative to null hypothesis testing based on $p$-values.  However, Bayesian model selection also has drawbacks.  Specifically, model selection from the Bayesian perspective is typically quite sensitive to prior choice on parameters, even ones deep within a hierarchy.  Scientific reporting can be improved if these sensitivities are more broadly recognized and easily reported among practicing statisticians and the wider research community. 

In the statistical literature, the sensitivity to hyperparameters (parameters to the priors on parameters) has been mentioned previously, see e.g. \cite{Berger1985}, \cite{Kass1995}, \cite{Berger1996}, \cite{Ohagan1995}, \cite{Chipman2001}, and \cite{Yao2018}. Yet, it remains difficult for practitioners who are interested in using Bayesian model selection in their research to understand the degree of sensitivity they might expect in their specific analysis plan. A researcher hopes their Bayesian model selection ``will be relatively insensitive to reasonable choices [of hyperparameters]'' \cite[][p. 146]{Berger1985} but worries that ``... such hypothesis tests can pose severe difficulties in the Bayesian inferential paradigm where the Bayes factors may exhibit undesirable properties unless parameter prior distributions are chosen with exquisite care'' \citep{Ormerod2017}. 

There is currently a dearth of practical strategies to help researchers systematically assess and report sensitivity in Bayesian model selection. In order for an approach to be useful for this purpose, it must (i) be simple to implement, not requiring extensive re-conceptualization in different scenarios, (ii) be efficient in instances where computing Bayes factors is computationally expensive, (iii) readily reveal the extent to which Bayes factors are sensitive to prior specification and facilitate the reporting of potential sensitivity.

With the above requirements in mind, we propose systematic assessment of the Bayes factor surface with the aid of modern libraries and visualization tools.  We define the Bayes factor surface as the value Bayes factors (usually on the log scale) take as a function of the hyperparameters in the model. Bayes factor surfaces are easy to visualize, require only the ability to form a Bayes factor in order to produce, and can be compared with available automatic procedures which appear as planes in Bayes factor surface visualizations.  We propose approximating Bayes factor surfaces using computer surrogate modeling \citep[e.g.,][]{sant:will:notz:2003} if computation of a given Bayes factor is expensive.  Importantly, Bayes factor surfaces inform their audience how inferences on competing models might vary as a function of prior belief about the parameters. We suspect that the instability revealed by Bayes factor surfaces might be considerably higher than many practicing statisticians realize, thus revealing a potential barrier for the greater adoption of Bayesian model selection in data analysis.

\begin{figure}[ht!] 
\centering
\includegraphics[width=180mm,trim=0 0 0 0]{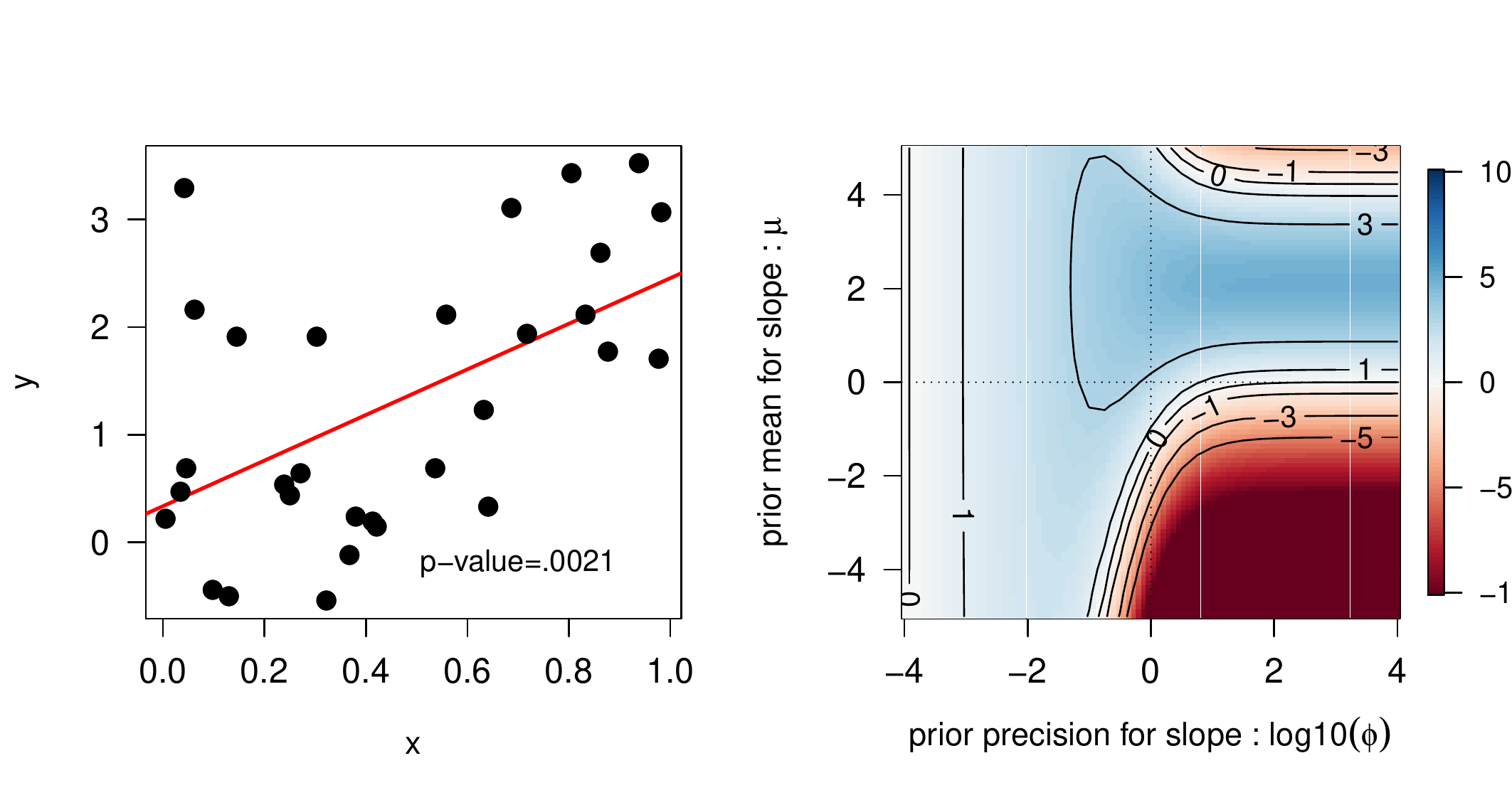}
\caption[]{Scatter plot (left panel) and Bayes factor surface contour plot (right panel).  Contour plot shows Bayes factor comparing $\beta=0$ and $\beta \neq0$ hypotheses on natural log scale.  Contours correspond to strength of evidence according to \cite{Kass1995}.   

}
\label{Fig1}
\end{figure}

Figure \ref{Fig1} shows the dilemma.  Consider a simple linear regression problem where the researcher wishes to test whether the slope parameter $\beta$ is equal to zero or not. The data in the left panel were simulated with $\beta=2.5$, error variance $\sigma^2=1$, $x$ values from a $\text{Uniform}(0,1)$ distribution, and sample size  $n=30$. The classical $p$-value for the slope is 0.0021 (least squares slope estimate of 2.12, standard error of 0.63, and $t_1=   3.38$), below the recently proposed $\alpha=0.005$ criterion to declare statistical significance for new discoveries \citep{Benjamin2018}. The contour plot shows the natural log of the Bayes factor (i.e., the Bayes factor surface) comparing $H_0:\beta=0$ and $H_1: \beta \neq0$ hypotheses. The $x$- and $y$-axes of the contour plot represent prior hyperparameters---user chosen inputs that encode prior belief. (We define these in detail in Section \ref{subsec:regression}.) Blue favors $H_1$ and red favors $H_0$. It is clear that the prior belief that the researcher brings to this analysis plays a major role in the conclusion. Decreasing prior precision ($x$-axis) yields equivocal evidence about hypotheses despite the apparent trend in the data.  Precise prior beliefs near the least squares estimate yield Bayes factors ``strongly'' in favor of a non-zero slope \citep[according to the scale of evidence proposed by ][]{Kass1995}.  Precise prior beliefs that do not conform to the known structure of these simulated data (i.e., red in lower right corner) provide very strong evidence in favor of $H_0$.

Figure \ref{Fig1} is, at first blush, rather discouraging. Least squares regression is among the simplest of data analytic approaches in any analyst's toolbox.  Even in the simplest of cases, it is clear that priors on parameters must be handled with care.  The good news is that within the class of linear models, several well-calibrated automatic approaches are available for Bayes factors, which we review later. In more complicated settings such as the problems described in Sections \ref{subsec:HLM} and \ref{subsec:surmod}, fully Bayesian automatic procedures for model selection are less readily available (although \cite{Fonseca2008} is a good starting point to develop an automatic procedure for Example \ref{subsec:surmod}). In these latter cases, the Bayes factor surface can play an important role in scientific reporting.

A thorough overview of Bayes factors can be found in \cite{Kass1995}.  A useful guide for implementation of Bayesian model selection is \cite{Chipman2001}. The remainder of this article is organized as follows.  Section \ref{sec:overview} reviews the general formulation of Bayesian model selection via Bayes factors and develops the idea of Bayes factor surfaces. Section \ref{sec:examples} covers three examples of varying complexity. The first example includes an interactive {\sf R} {\tt shiny} \citep{Rshiny2017} application hosted at http://shiny.stat.vt.edu/BFSApp/ that allows the user to explore the Bayes factor surface in real time.  The second example employs the Bayes factor surface approach to determine which hyperparameters induce sensitivity for hierarchical linear model selection. The third example addresses the common situation where Bayes factor approximation is computationally intensive, and often a ``noisy'' enterprise owing to the stochastic nature of approximation via Markov chain Monte Carlo (MCMC). Therein we propose a space-filling design in the hyperparameter space, and surrogate modeling with Gaussian process models to emulate the Bayes factor surface.  Modern libraries in {\sf R}, like {\tt hetGP} \citep{hetGP} on CRAN, cope well with both mean and variance features common in Bayes factor surfaces.  Our fully reproducible illustration, with code provided in the supplementary material, is designed to be easily ported to Bayes factor calculations in novel contexts.  Discussion is included in Section \ref{sec:discussion}.

\section{Bayesian model selection overview}
\label{sec:overview}

Let $\bm{y}$ represent observed data.  Let $M_1$ and $M_2$ represent two competing models (or hypotheses), and let $\bm{\theta}_1$ and $\bm{\theta}_2$ represent vectors of parameters associated with each of $M_1$ and $M_2$, respectively. Index models and parameter vectors with $k=1,2$. Let the distribution of the data given the parameters (i.e., the likelihood normalized as a proper distribution) be denoted $p(\bm{y}|\bm{\theta}_k,M_k)$. 

The marginal likelihood for model $k$ (denoted $p(\bm{y}|M_k)$) arises when parameters are integrated out of the joint distribution of the parameters and data, i.e.,
\begin{equation}
\label{marglike}
p(\bm{y}|M_k)=\int p(\bm{y}|\bm{\theta}_k,M_k) p(\bm{\theta}_k |M_k) \; d \bm{\theta}_k.
\end{equation}
The marginal likelihood plays a central role in Bayesian model comparison, selection, and averaging. The Bayes factor is the usual metric to compare $M_1$ and $M_2$ and is the ratio of those models' marginal likelihoods:
\begin{equation}
\mathrm{BF}_{12} = \frac{p(\bm{y}|M_1)}{p(\bm{y}|M_2)}.
\label{BF}
\end{equation}

In this work we consider Bayes factors based on parametric Bayesian models.  Historically,  scales of interpretation \citep{Jeffreys1961,Kass1995} have been used to describe strength of evidence in favor of $M_1$ over $M_2$ where large values of $\mathrm{BF}_{12}$ suggest that observed data support $M_1$ more than $M_2$.  However, using scales in this manner implicitly assumes prior model probabilities $p(M_1)=p(M_2)=\frac{1}{2}$, which is the unique setting where the Bayes factor describes the ratio of how much more probable $M_1$ is compared with $M_2$. In general, interpreting Bayes factors as the weight of evidence in favor of one hypothesis over another is not strictly justified. Rather, Bayes factors summarize the extent to which the observed data update prior model odds to posterior model odds. Henceforth we assume equal model priors which justifies the interpretation of Bayes factors in terms of relative probability of competing models.  The reader can adopt other model priors and interpret the Bayes factor as a multiplicative factor which updates prior belief about models. For a more complete discussion of this and related issues, see \cite{Lavine1999}.

The notation $p(\bm{y}|M_k)$ for marginal likelihoods shown in Equation (\ref{marglike}) is remarkably consistent across the statistical literature.  Perhaps for brevity, this notation omits reference to the parameters that have been integrated, which facilitates an illusion that $p(\bm{y}|M_k)$ does not depend on the priors assigned to $\bm{\theta}_k$. In the remainder of this paper we show that choices about the priors on $\bm{\theta}_k$ can exert strong influence on Bayes factors.

The first famous discussion about the sensitivity to priors for Bayesian model selection arose as the Jeffreys-Lindley-Bartlett paradox \citep{Jeffreys1935,Jeffreys1961,Bartlett1957,Lindley1957}. The issue (which is not technically a paradox) was described \citep{Lindley1957} in terms of a one parameter point null testing problem in which frequentist $p$-values strongly favor rejecting the null while Bayes factors provide strong support for the null. This simple example illustrates that as the prior on the parameter being tested becomes vague, the Bayes factor increasingly favors the simpler point null hypothesis. See \cite{Ormerod2017} for thorough discussion and numerous references to the ``paradox'' in the literature. Bayes factor surfaces allow for the visualization of the effects of prior vagueness and other hyperparameters simultaneously.

The methods described thus far apply when only two models are under consideration.  When $K>2$ models are of interest, usually a series of $K-1$ Bayes factors are produced, where a specific ``base'' model is chosen, and a Bayes factor is formed comparing all other models to the base model.  In cases such as these, the researcher may wish to examine Bayes factor surfaces corresponding to each of the $K-1$ Bayes factors that are considered.

\subsection{Bayes factor surfaces}
The method we propose in this paper is simple.  We suggest producing a graphical display of the log Bayes factor as a function of  hyperparameters of interest for inclusion in scientific reports. Basically, enumerate a grid in the hyperparameter space, collect Bayes factors under each setting, and stretch over an appropriate mesh---off-loading the heavy visual lifting, projections, slices, etc., to any number of existing rich visualization suites.  One drawback to this approach is that numerical integration to obtain marginal likelihoods (\ref{marglike}) is frequently expensive.  Even when closed-form expressions are available, their evaluation on a mesh for visualization can be computationally prohibitive, especially when the hyperparameter space is large.   When Monte Carlo integration is used the output can be noisy.  Although that noise can be reduced with further computation (i.e., longer MCMC chains), convergence is typically not uniform in the hyperparameter space.   For a fixed simulation effort in all hyperparameter settings, the level of noise in the output Bayes factor may change.  In other words, the Bayes factor response surface can be heteroskedastic. In situations such as these, we propose treating Bayes factor calculation as a computer simulation experiment: limited space-filling design in the hyperparameter space and surrogate modeling via Gaussian processes in lieu of exhaustive evaluation and inspection.  We illustrate a modern library called {\tt hetGP} which copes with input dependent variance and leads to nice surrogate Bayes factor surface visualizations in our examples.

The simplicity of the approach is a virtue. The core idea does not need to be reconceptualized in different settings and could be explored in any situation where Bayes factors are available and concern about sensitivity is present. For example, a Bayes factor surface plot could accompany statistical reporting as a method to indicate to reviewers and the broader audience whether the conclusions with respect to model selection are driven heavily by the specific priors chosen for $\bm{\theta}_k$. As a corollary this approach also fosters transparency since the audience can see how model selection would differ under other hyperparameter settings. 

\section{Examples}
\label{sec:examples}

\subsection{Ordinary regression}
\label{subsec:regression}

The first example comes from the ordinary regression setting with model

\begin{equation}
y_i=\alpha + \beta x_i + \epsilon_i
\label{eq:SLR}
\end{equation}
where $\alpha$ is the $y$-intercept, $\beta$ is the slope, $i=1,\ldots,n$, $x_i$ represents fixed known data from a predictor variable, and $y_i$ is the outcome. We assume $\epsilon_i \stackrel{\mathrm{iid}}{\sim}\mathcal{N}\left(0,\frac{1}{\gamma}\right)$, where the error precision is $\gamma= \frac{1}{\sigma^2}$ and $\sigma^2$ is the usual error variance. Specifying error in terms of precision is mathematically convenient for Bayesian analysis. Following Equation (\ref{BF}) and the color scale in Figure \ref{Fig1}, under $M_1$ we have $\beta \neq 0$ and under $M_2$  $\beta = 0$.  The remaining model specifications for $M_1$ and $M_2$ share the following features:
\begin{align}
\label{eq:priorcoef}
\beta|M_1 &\sim \text{Normal}(\mu, \phi) \\
\label{eq:priorcoef0}
\beta|M_2 &=0 \\
\label{eq:priorgamma}
\gamma &\sim \text{Gamma}(a,b) \\
\label{eq:priorint}
p(\alpha) &\propto 1 \\ 
\label{eq:liklihood}
y_i|\alpha, \beta, \gamma, M_k &\stackrel{\mathrm{iid}}{\sim} \text{Normal}\left(\alpha+\beta x_i, \frac{1}{\gamma} \right).
\end{align}
  
The parameters in each model are vectors $\bm{\theta}_1=(\alpha,\beta, \gamma)^\top$ and $\bm{\theta}_2=(\alpha,\gamma)^\top$. It has been shown previously that putting a flat prior on the intercept implicitly centers the $y_i$ and $x_i$ about their respective means during the computation of the marginal likelihood \citep{Chipman2001}. Denote $w_i = x_i-\bar{x}$ and $z_i = y_i-\bar{y}$.

Under the simpler hypothesis $M_2:\beta=0$ and (\ref{eq:priorgamma}--\ref{eq:liklihood}),  the marginal likelihood has a convenient closed-form solution.
\begin{equation}
\label{margnull}
p(\bm{y}|M_2)=  \frac{c b^a (2 \pi)^{\frac{-(n-1)}{2}} \Gamma\left(\frac{n-1}{2}+a\right) }{\Gamma(a) \sqrt{n} \left[b + \frac{1}{2}\sum_{i=1}^{n} z_i  ^2 \right]^{\frac{n-1}{2}+a}},
\end{equation}
 where $c$ is an undetermined constant that arises from the improper prior (\ref{eq:priorint}).  As noted in Section \ref{sec:overview}, the standard notation for marginal likelihoods, as in Eq.~(\ref{margnull}),  omits reference to $\bm{\theta}_2$ due to its integration.  Yet, obviously specific choices about hyperparameters ($a$ and $b$ in this case) associated with $\bm{\theta}_2$ appear in the marginal likelihood. The form of prior densities also influences the marginal likelihood in a less visible manner. Even though standard notation tends to obscure this fact, prior choice on parameters directly influences marginal likelihoods and thus Bayesian model selection. Across the universe of possible Bayesian model selection endeavors, there are no ironclad guarantees that increasing $n$ diminishes the influence of hyperparameters.

Under the $M_1:\beta \neq 0$, integration of $\alpha$ and $\beta$ can be handled analytically, but integration of $\gamma$ requires numerical approximation.
\begin{align}
p(\bm{y}|M_1) &= \frac{c(2\pi)^{\frac{-(n-1)}{2}} b^a}{\Gamma(a) \sqrt{n}} \bigintsss \frac{\sqrt{\phi} \gamma^{\frac{n-1}{2}+a-1}}{\left(\phi + \gamma \sum_{i=1}^n w_i^2 \right)^{\frac{1}{2}}} \label{margalt} \\
 &\quad \times \exp\left(-b\gamma -\frac{1}{2}\left(\phi\mu^2 + \gamma\sum_{i=1}^{n}z_i^2\right) 
 + \frac{1}{2} \frac{\left(\phi\mu + \gamma \sum_{i=1}^{n} z_i w_i\right)^2}{(\phi + \gamma \sum_{i=1}^n w_i^2)}\right) \; d\gamma \nonumber.
\end{align}

The marginal likelihoods (\ref{margnull}--\ref{margalt}) in this example can be obtained by exploiting recognizable exponential family kernels to perform the integration in (\ref{marglike}), although the one-dimensional integral over $\gamma$ in (\ref{margalt}) requires numerical approximation.

Choice of hyperparameters $a$, $b$, $\phi$, and $\mu$ enable the researcher to incorporate prior information about parameters into the analysis.  In the absence of specific prior information, the priors on $\beta$ and $\gamma$ are frequently taken to be vague yet proper.  For example, a researcher may center $\beta$ at zero and choose a small prior precision (i.e., large prior variance) and similarly for $\gamma$ they may choose to center at 1 and make the prior variance large.  This is an attempt to impart minimal information to the inference from the prior. The improper flat intercept $\alpha$ contributes an unspecified constant to the marginal likelihood in each model which cancels once the Bayes factor is formed.  

The included {\sf R} {\tt shiny} \citep{Rshiny2017} application here http://shiny.stat.vt.edu/BFSApp/ allows users to entertain various settings of hyperprior, sample size, and true underlying effect size $\beta$ in order to explore how these affect the Bayes factor surface. Experimentation with this application illustrates that hyperparameters for parameters common to both models (e.g., $a$ and $b$ specified for $\gamma$) tend to have much less influence on Bayes factors than hyperparameters for parameters that are being tested (e.g., $\mu$ and $\phi$ specified for $\beta$).

Of course, adopting the priors (\ref{eq:priorcoef}-\ref{eq:priorint}) is only one possible strategy in a universe of available Bayesian priors.  In this study we will also consider three specific automatic procedures to obtain Bayes factors that are readily available for regression. The term ``automatic'' is used in Bayesian procedures to describe cases where the researcher does not have to specify priors subjectively. We include Zellner--Siow mixture $g$-priors \citep{Liang2008}, Bayes factor approximation via Bayesian information Criterion 
\citep[][see also \citealt{Kass1995}]{Schwarz1978}, and fractional Bayes factors \citep{Ohagan1995}, each of which is briefly summarized next. As the reader probably expects, these automatic procedures are each  defensible individually, yet do not necessarily agree and could potentially lead to different conclusions in different situations.

{\bf Zellner--Siow mixture $g$-prior:}
The Zellner--Siow mixture $g$-prior was introduced \citep{Liang2008} as an automatic Bayesian specification to allow selection among nested ordinary linear models. Previous work \citep{Zellner1980} has shown that a multivariate Cauchy distribution on regression coefficients satisfies basic model consistency requirements.  \cite{Liang2008} demonstrates that an inverse gamma with shape and rate parameters of $\frac{1}{2}$ and $\frac{n}{2}$ respectively, placed on the $g$ parameter of Zellner's $g$-prior \citep{Zellner1986} induces Cauchy tails on the prior of $\beta$ while only requiring a one-dimensional approximation to the integral for the marginal likelihood.  Appendix \ref{app:regress} includes a brief description of the Bayesian model for the Zellner--Siow mixture $g$-prior.

{\bf Approximation by BIC:}
Asymptotically,
$\mathrm{BF}_{12} \approx  e^{-\frac{1}{2}\left(\mathrm{BIC}_1-\mathrm{BIC}_2 \right)}$
where $\mathrm{BIC}_k$ is the usual Bayesian information criterion \citep{Schwarz1978} for model $k$. Further details relating BIC to approximate Bayesian model selection can be found in \cite{Kass1995}.

{\bf Fractional Bayes factor:}
The final automatic method we consider is the fractional Bayes factor \citep{Ohagan1995}.  Fractional Bayes factors are a variant of partial Bayes factors \citep{Berger1996} which reserve a training sample from the full data, update the prior using the training sample, then form a Bayes factor using the remainder of the data in the likelihood.  A central appeal of the partial Bayes factor approach is that parameters with noninformative priors (which are frequently improper) can be tested in an essentially automatic fashion.  While intrinsic Bayes factors \citep{Berger1996} are obtained by averaging over many partial Bayes factors, fractional Bayes factors obviate the need to choose a specific training sample through a clever approximation of the ``partial'' likelihood and are thus computationally less expensive than intrinsic Bayes factors. In this approach we adopt noninformative improper priors and compute a Bayes factor surface using fractional Bayes factors. More detail can be found in Appendix \ref{app:regress}.

\subsubsection*{{\sf R} {\tt shiny} application}

To allow the reader to interactively explore Bayes factor surfaces in the regression setting, we have produced an R shiny application \citep{Rshiny2017} located here \\ http://shiny.stat.vt.edu/BFSApp/. The application allows the user to set sample size and parameter values to simulate regression data, and also to choose hyperparameters $a$, $b$, $\phi$, and $\mu$. The application produces the contour plot in Figure \ref{Fig1}, a three-dimensional rotating Bayes factor surface plot with options to superimpose planes corresponding to each of the three automatic procedures described in Section \ref{subsec:regression}, and density plots for the priors. The purpose of this application is to allow the reader to generate data with a known regression structure, then interactively examine sensitivities in the Bayes Factor surface.  Source and readme files and an instructional video are included in the supplementary files for this paper.

\subsection{Hierarchical linear models}
\label{subsec:HLM}

\begin{figure}[ht!] 
\centering
\includegraphics[width=70mm,trim=0 0 0 0]{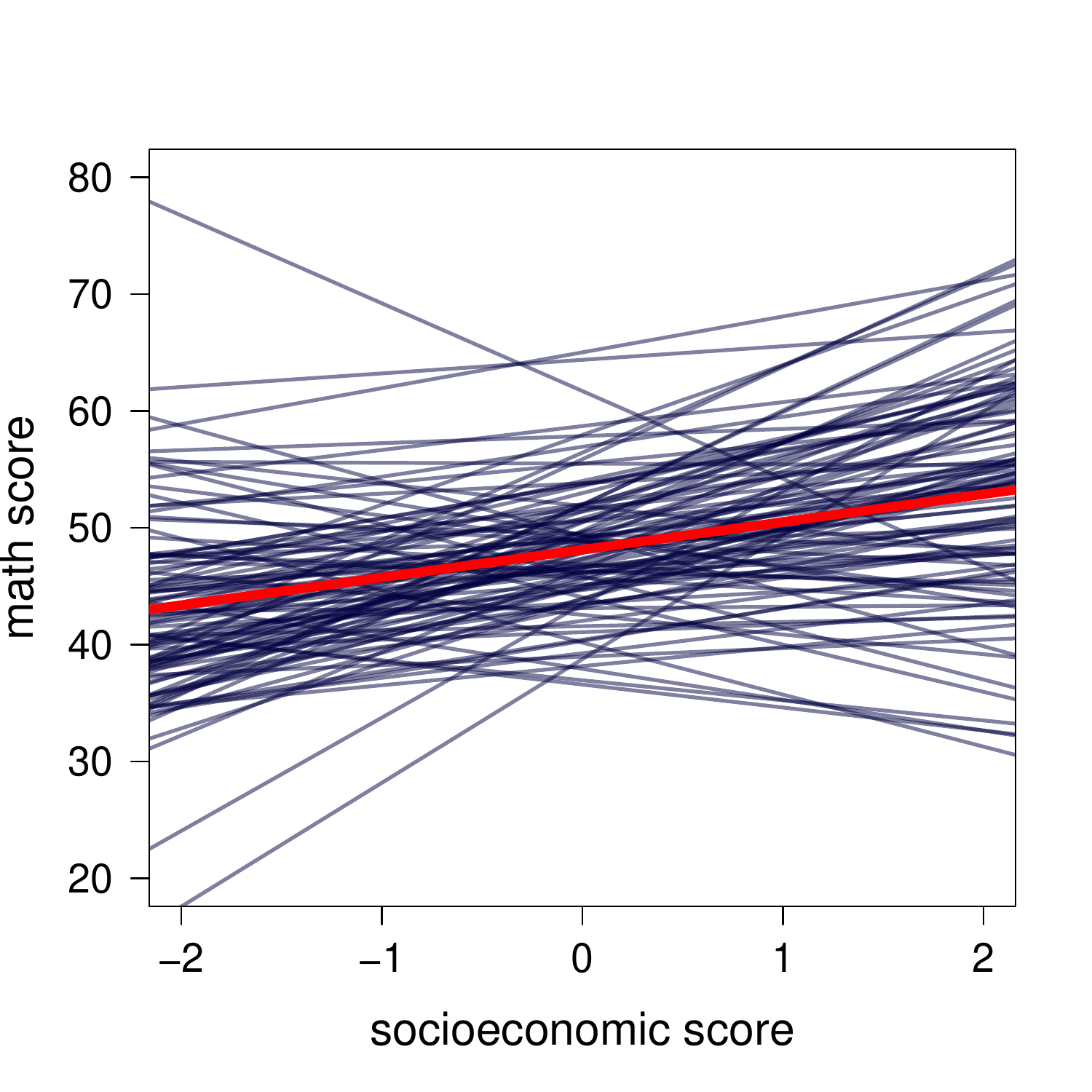}
\caption[]{Least squares fits for the math score data presented in \cite{Hoff2009}.  Each blue line shows the estimated regression line of math score on socioeconomic status within a school. The red line's slope and y-intercept are the averages of the estimated slopes and y-intercepts for the schools.}
\label{schooldatsum}
\end{figure}

We next consider competing hierarchical linear models for student math scores \citep[][Chapters 8 \& 11]{Hoff2009}. These data include $1,993$ students in $m=100$ schools, where math scores are the outcome variable. We assess socieoeconomic score (centered within each school) as a candidate predictor. Least squares fits to these data are presented in Figure \ref{schooldatsum}.

Figure \ref{schooldatsum} suggests (i) that there is substantial variability in school-level effects, and (ii) a potential overall effect for socioeconomic score on math score. The Bayes factor surface approach allows us to determine sensitivity in model selection as a function of hyperparameter specification related to (ii) while honoring both student- and school-level variability described in (i). For model $M_1$ we consider a ``slopes and intercepts'' model that accounts for the school-specific linear effect of socioeconomic status on math score.  For $M_2$ we consider a simpler ``means'' model for the school effects in which schools have variability in average math score which is not related to socioeconomic score. 

To examine the impact of hyperparameter specification on Bayes factors in this problem, we modify the original hierarchical models posed in \cite{Hoff2009}.  We impose a fixed $g$-prior on school-level effects to make marginal likelihood computations more tractable. Both $M_1$ and $M_2$ are special cases of the following model: 
\begin{align}
y_j|\beta_j,\gamma,M_k & \sim \text{Multivariate Normal}_{n_j} \Big(X_j \beta_j,\frac{1}{\gamma}I_{n_j}\Big) \nonumber \\
\beta_j|\theta, \gamma,M_k & \sim \text{Multivariate Normal}_{p_k} \Big(\theta,\frac{g}{\gamma}(X_j^\top X_j)^{-1}\Big) 
\label{eq:HLMod}\\
\theta|M_k & \sim \text{Multivariate Normal}_{p_k}\Big(\mu_0,\Lambda_0 \Big) \nonumber \\
\gamma & \sim \text{Gamma}\Big(\frac{\nu_0}{2},\frac{\nu_0 \sigma_0^2}{2}  \Big) \nonumber
\end{align}
where $j=1,\ldots,100=m$ indexes schools, $y_j$ is a length $n_j$ vector that contains the $j^{\mathrm{th}}$ school's math scores, $n_j$ is the number of student data points in the $j^{\mathrm{th}}$ school, $\beta_j$ is a vector of regression effects in the $j^{\mathrm{th}}$ school, $X_j$ is a $n_j \times p_k$ model matrix for the $j^{\mathrm{th}}$ school, $\gamma$ is error precision, and $\theta$ is the prior mean on the regression coefficients. User-selected hyperparameters include $g$, $\mu_0$, $\Lambda_0$, $\nu_0$, and $\sigma^2_0$. Details on the computation of the $p(\bm{y}|M_k)$ marginal likelihoods needed to compute the Bayes factor are in Appendix \ref{app:hlm}.

As is often the case for hierarchical models, $M_1$ and $M_2$ have a relatively large number of random effects (200 and 100, respectively), but these random effects are governed by a ``between-group'' distribution with relatively few parameters.  Bayes factor surfaces can be used to examine the effect of hyperparameter specification of the between-group model, and are thus useful in the hierarchical modeling context.

Note that unlike the mixture $g$-prior formulation in Example \ref{subsec:regression}, we employ a fixed $g$ specification, then vary $g$ as a hyperparameter on the Bayes factor surface. Within the variable subset selection literature that gave rise to the $g$ prior, fixed $g$ specifications have been largely abandoned in favor of mixture $g$ priors, since this latter set has desirable model selection consistency properties within the class of ordinary regression \citep{Liang2008}. Our adoption of a fixed value for $g$ is meant to illustrate (i) the potential hazards of using a fixed value of $g$ in the hierarchical model selection scenario, (ii) the ability of the Bayes factor surface to identify sensitivity in the specification of a hyperparameter such as $g$ when many other hyperparameters are also included in the candidate models, and (iii) because there is currently a dearth of theoretical results for mixture $g$-prior-based model selection in the literature for hierarchical models.

\begin{figure}[ht!] 
\centering
\includegraphics[width=180mm,trim=0 0 0 0]{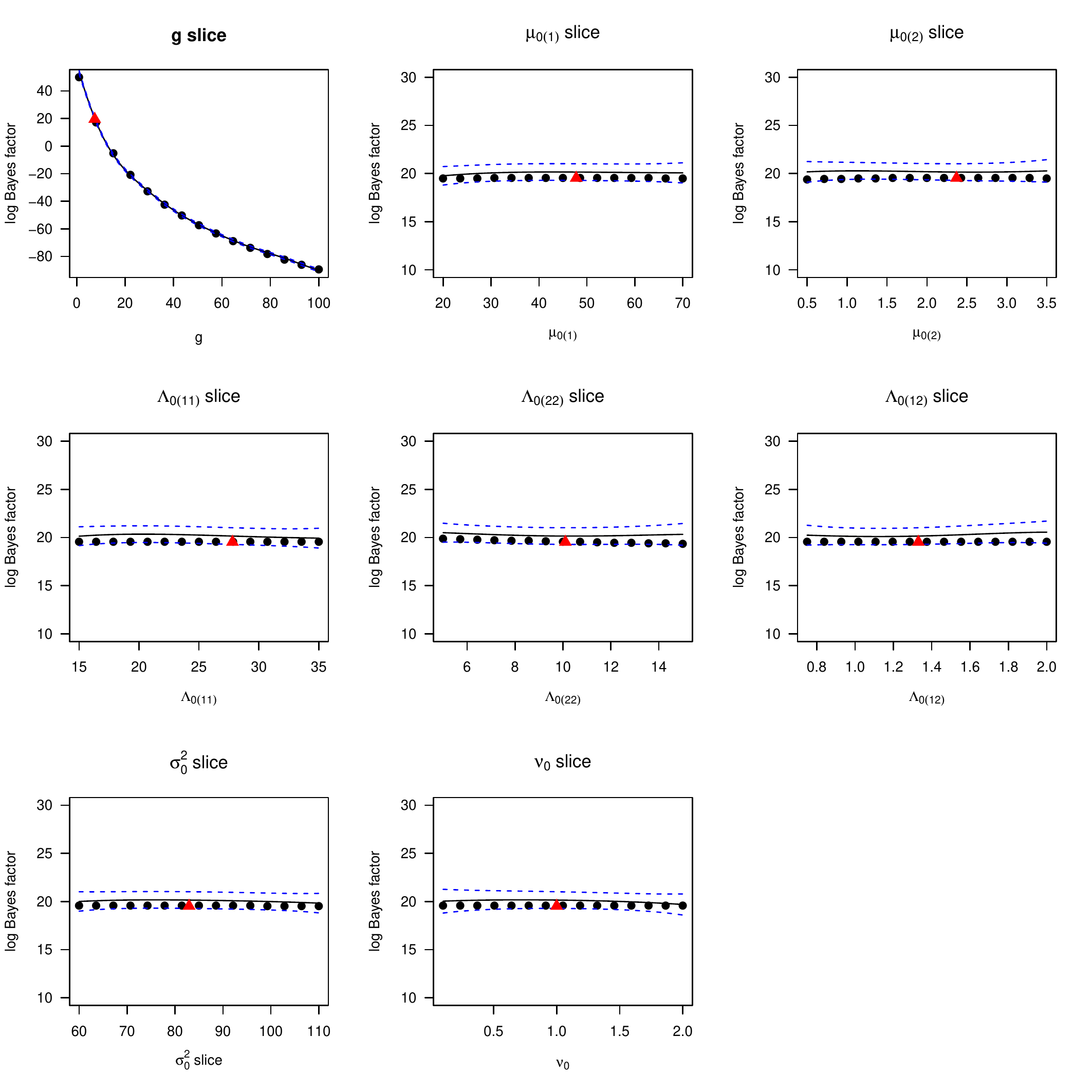}
\caption[]{Bayes factor surface slices as a function of hyperparameters. Large values lend support to ``slopes and intercepts'' model $M_1$, and small values support ``means'' model $M_2$. Black lines are Gaussian process fits with corresponding 90\% error bars as blue dotted lines. Black circles represent grid points on which Bayes factor was evaluated.  Red triangles represent ``default'' values of hyperparameters (see text). From top left to bottom right, hyperparameters are $g$, prior mean on y-intercept, prior mean on slope, prior variance on y-intercept, prior variance on slope, prior covariance between y-intercept and slope, and hyperparameters for error precision.}
\label{FigHLMBF}
\end{figure}

Figure \ref{FigHLMBF} shows one-dimensional slices of the Bayes factor surface for each of the eight hyperparameters in this study. Each of these slices varies the value of one hyperparameter. Dots show 15 realized values of the Bayes factor surface and fitted lines and confidence bands arising from a Gaussian process fit are the subject of a discussion in Section \ref{subsec:surmod}. Triangles show data-inspired ``default'' levels and indicate the value of the hyperparameters where the other slices are visualized. These default values follow the reasoning in \cite{Hoff2009} and are:
\begin{align*}
\mu_0&=(47.76,2.37)^\top, &\Lambda_0&=\begin{pmatrix} 
27.80 & 1.33 \\
1.33 & 10.10 
\end{pmatrix}, &\sigma^2_0&=82.94 & \nu_0&=1 & \mbox{and} \quad g&=7.44. 
\end{align*}
This value of $g$ was obtained by setting the empirical covariance matrix based on the 100 school-level estimates of the least squares estimates of the $\beta_j$ values equal to the average of $\frac{g}{\gamma}(X_j^\top X_j)^{-1}$ for $M_2$. (Applying the same strategy, $g$ is between 7 and 8 based on the ``slopes and intercept'' model $M_1$). 

This investigation reveals the extraordinary extent to which $g$ governs the result of model selection, apparent from the top left panel of Figure \ref{FigHLMBF}. Interestingly, recommended values for fixing $g$ based on ordinary regression (which predate mixture $g$ prior methodology and were not devised with hierarchical modeling in mind) include setting $g$ equal to the sample size \citep{KassWasserman1995}, the squared number of parameters \citep{Foster1994}, and the maximum of sample size and squared parameters \citep{Fernandez2001}. If a researcher hastily adopted any of these choices they would favor the simpler ``means'' model $M_2$. However, our empirical strategy for $g$, which captures the value of $g$ most consistent with the data, clearly favors $M_1$.  Also, BIC-based comparison of ``slopes and intercepts'' ($\text{BIC}=14,637$) and ``means'' ($\text{BIC}=14,707$) models \citep[using the {\tt lme4} package in R,][]{Bates2015}, favors the ``slopes and intercepts'' model which corresponds to our $M_1$ (BIC does not assume precisely the model we propose). The BIC analysis agrees with the empirical $g$ approach (regardless of whether $g$ is learned from $M_1$ or $M_2$), and corroborates the apparent overall trend that socioeconomic score shares with math score in Figure \ref{schooldatsum}. The rest of the hyperparameters behave consistently with the principles discussed in Example \ref{subsec:regression}, and none of them affect the analysis to nearly the extent that $g$ does. A brief comment on the mathematical intuition for why $g$ induces such sensitivity in the Bayes factor for this example can be found in Appendix \ref{app:hlm}.

Even though we began this exercise with optimism that (\ref{eq:HLMod}) would yield fruitful model selection, we are forced to conclude that the proposed model is overly sensitive to the hyperparameter $g$. Fortunately, the Bayes factor surface approach makes this particular sensitivity easy to detect.  Without the use of Bayes factor surfaces, researchers might not have good intuition for which hyperparameter(s) induce sensitivity into model selection, or even know whether a given Bayesian model formulation has the potential to offer reliable model selection.

\subsection{Surrogate modeling of the Bayes factor surface}
\label{subsec:surmod}

In the simple regression setting of Section \ref{subsec:regression}, computation is cheap and many viable automatic procedures are available.  In many other settings, obtaining marginal likelihoods is a delicate and computationally expensive task.  For cases such as these, we propose that an adequate analogue of the visualizations provided above can be facilitated by emulating the Bayes factor surface in the hyperparameter space of interest.  In other words, we propose treating expensive Bayes factor calculations as a computer simulation experiment. Evaluations at a space-filling design in the (input) hyperparameter space can be used to map out the space and better understand the sensitivity of model selection to those settings.

In Section \ref{subsec:HLM} the hyperparameter space was high-dimensional.  Although Bayes factors may be evaluated in closed form, each calculation by no means represents a trivial expense. Exhaustively gridding out such a space to any reasonable resolution, and running Bayes factor calculations on each element of the grid, is intractable.  Obtaining evaluations along slices, as in Figure \ref{FigHLMBF}, requiring $5 \times 8 = 120$ evaluations is manageable, taking about eight minutes of compute time.  A $10^8$ grid would take thousands of days by contrast.  Instead, evaluating on a space-filling Latin hypercube design \citep{mcka:cono:beck:1979} of size $n=1000$ is manageable in about an hour.  Off-the-shelf Gaussian process surrogate modeling software -- see supplementary material -- can be used to develop a surrogate for these runs.  An example is shown in terms of predictive means and quantiles overlayed as lines in Figure \ref{FigHLMBF}.  As can be seen, coverage is quite good: 96.6\% compared to the nominal 95\% in this case.  The surrogate may be used to investigate other features of the surface that might be of interest, such as main effects or sensitivity indices.  

In the context of computer surrogate modeling, the Bayes factor ``simulator'' from Section \ref{subsec:HLM} represents a rather tame exemplar, yielding a highly smooth surface, relatively flat in most inputs (hyperparameters), and noise free.  Consequently, we don't dwell on many of the details here.  The case where Bayes factor evaluation is not available in closed form, but instead requires numerical evaluation (e.g., with MCMC), is rather more interesting.  In this case, more sophisticated machinery is necessary.  

As motivation, consider an experiment described by \citet[][Section 3.3--3.4]{Gramacy2010} involving Bayes factor calculations to determine if data are leptokurtic (Student-$t$ errors) or not (simply Gaussian).  Some context for that experiment is provided here with the primary goal of conveying that the setup is complicated, involving the confluence of many methodological and computational tools.  The setting is portfolio balancing with historical asset return data of varying length.  Highly unbalanced history lengths necessitated modeling in Cholesky-space, through a cascade of Bayesian linear regressions decomposing a joint multivariate normal via conditionals.  Modern regularization priors including double-exponential (lasso), Gaussian (ridge), normal-gamma (NG), and horseshoe under augmented with reversible-jump (RJ) scheme for variable selection were considered.  Their {\tt monomvn} package \citep{monomvn} on CRAN provided RJ-MCMC inference for the regression subroutines, and subsequently for reconstruction of the resulting joint MVN distributions of the asset returns \citep{andersen:1957,stambaugh:1997}, which are funneled through quadratic programs for \citet{markowitz:1959}-style portfolio balancing.  With the proposed methods coming online just after the 2008 financial crisis, the authors were particularly interested in the appropriateness of the prevailing MVN modeling setup---theirs being one example---in the presence of data which may contain outliers, or otherwise heavy tails.  Comparing different models for tail behavior given the observed data is where the present approach becomes relevant.

To accommodate robust inference in the face of outliers, \citeauthor{Gramacy2010} tacked the latent-variable-based Student-$t$ sampler from 
\citet{geweke:1992,geweke:1993} onto their inferential framework.  To determine if the data warranted heavy-tailed modeling, they further augmented the software with  a Monte Carlo Bayes factor calculation ($\mathrm{BF}_{\mathcal{N}\mathrm{St}}$) in the style \citet[][Section 2.5.1]{jacq:polson:rossi:2004}, leveraging that Student-$t$ (St) and normal ($\mathcal{N}$) models differ by just one parameter in the likelihood: $\nu$, the degrees of freedom.  A subsequent simulation study, however, casts doubt on whether or not one could reliably detect heavy tails (when indeed they were present) in situations where the generating $\nu$-value was of moderate size, say $\nu \in \{3,4,5,6,7\}$.  It is relevant to ask to what extent their conclusions are impacted by their hyperparameter choices, particularly on the prior for $\nu$ which they took to be $\nu \sim \mathrm{Exp}(\theta = 0.1)$.  Their intention was to be diffuse, but ultimately they lacked an appropriate framework for studying sensitivity to this choice.

Before a thorough analysis, first consider a simpler experiment. We set up a grid of hyperparameter values in $\theta$, evenly spaced in $\log_{10}$ space from $10^{-3}$ to 10 spanning ``solidly Student-$t$'' (even Cauchy) to ``essentially Gaussian'' in terms of the mean of the prior over $\nu.$  For each $\theta_i$ on the grid we ran the RJ-MCMC to approximate $\mathrm{BF}_{\mathcal{N}\mathrm{St}}$.  The training data were generated following a setup identical to the one described in \citet{Gramacy2010}, involving 200 observations randomly drawn from a linear model in seven input dimensions, with three of the coefficients being zero, and $\nu=5$.  Details are provided by our fully reproducible R implementation provided in the supplementary material.  In order to understand the Monte Carlo variability in those calculations, ten replicates of the BFs under each hyperparameter setting were collected.

\begin{figure}[ht!]
\centering
\includegraphics[scale=0.5,trim=0 0 0 55]{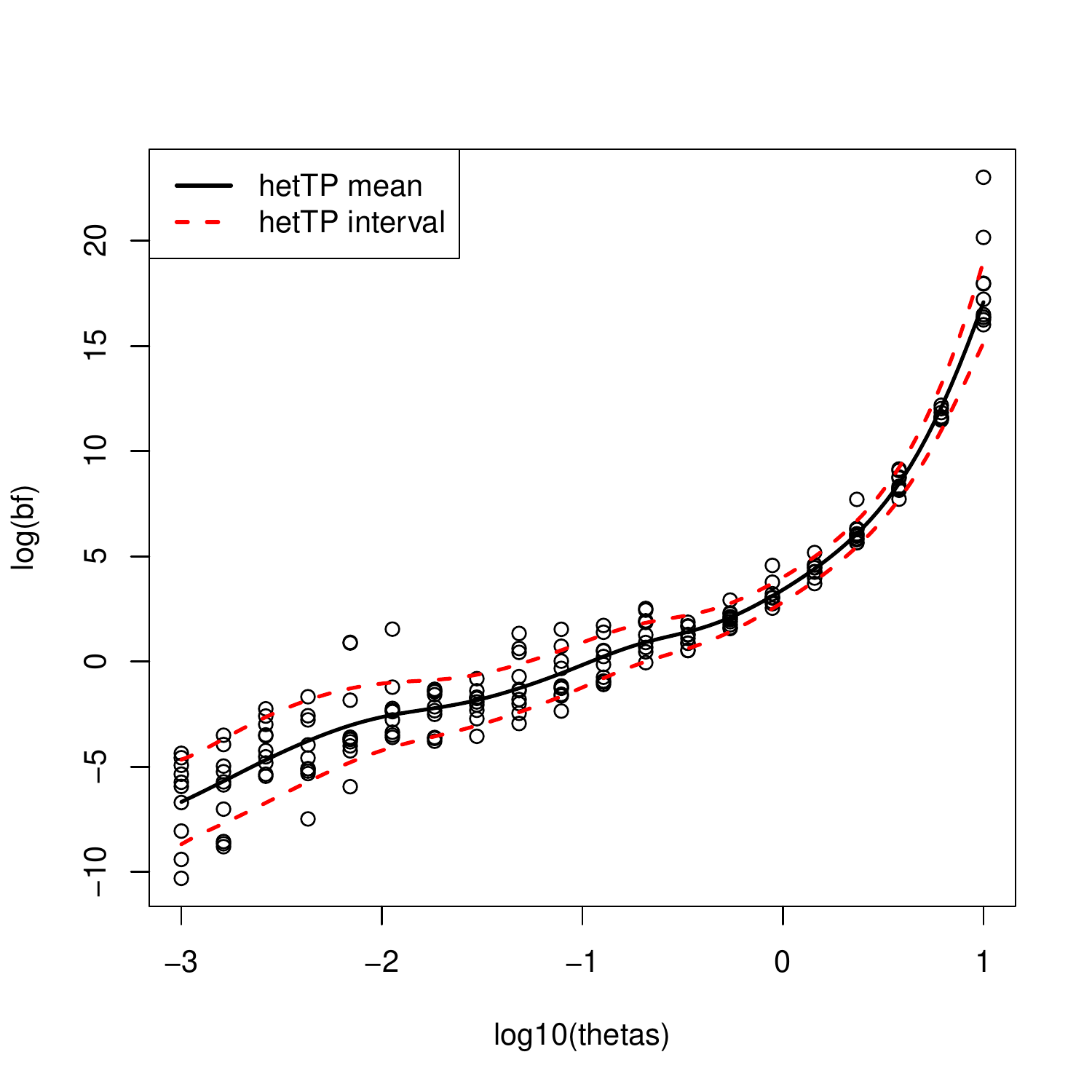}
\caption{Log Bayes factors for varying hyperparameter prior settings $\theta$ in  $\nu \sim \mathrm{Exp}(\theta)$.  The $x$-axis ($\theta$) is shown in $\log_{10}$ space.  A predictive surface from {\tt hetTP} is overlayed via mean and 95\% error-bars.}
\label{f:expbfs}
\end{figure}

The results are shown in Figure \ref{f:expbfs}.  Each open circle is a $\mathrm{BF}_{\mathcal{N}\mathrm{St}}$ evaluation, plotted in $\log_{10}-\log_e$ space.  Observe that outputs on the $y$-axis span the full range  of  evidence classes \citep{Kass1995} for the hypothesis of Student-$t$ errors over the Gaussian alternative.  When $\theta$ is small, the Student-$t$ is essentially a foregone conclusion; whereas if $\theta$ is large the Gaussian is.  The data, via likelihood through to the posterior, is providing very little in the form of adjudication between these alternatives.  An explanation is that the model is so flexible, the posterior is happy to attribute residual variability---what we know to be outliers from heavy tails (because we generated the data)---to mean variability via settings of the linear regression coefficients, $\beta$.  A seemingly innocuous hyperparameter setting is essentially determining the outcome of a model selection enterprise.

Each $\mathrm{BF}_{\mathcal{N}\mathrm{St}}$ evaluation, utilizing $T=100000$ MCMC samples, takes about 36 minutes to obtain on a 4.2 GHz Intel Core i7 processor, leading to a total runtime of about 120 hours to collect all 200 values used in Figure \ref{f:expbfs}.  Although that burden is tolerable (and perhaps we could have made do with fewer evaluations), extending to higher dimensions is problematic.  Suppose we wanted to entertain $\nu \sim \mathrm{Gamma}(\alpha, \beta)$, where the $\alpha=1$ case reduces to  $\nu \sim \mathrm{Exp}(\beta \equiv \theta)$ above.  If we tried to have a similarly dense grid, the runtime would balloon to 100 days, which is clearly unreasonable.  Rather, we propose treating the $\mathrm{BF}_{\mathcal{N}\mathrm{St}}$ calculation as a computer experiment: build a surrogate model from a more limited space-filling design, and use the resulting posterior predictive surface to understand variability in Bayes factors in the hyperparameter space.  Before providing an example, the simpler experiment in Figure \ref{f:expbfs} points to some potential challenges.  The $\mathrm{BF}_{\mathcal{N}\mathrm{St}}$ surface is heteroskedastic, even after log transform, and may itself have heavy tails.  

Fortunately, stochastic computer simulations with heteroskedastic errors are appearing throughout the literature and some good libraries have recently been developed to cope.  The {\tt hetGP} package on CRAN \citep{hetGP}, based on the work of \citet{Binois2018}  with Student-$t$ extensions \citep{Chung2018}, is easy to deploy in our context.  The predictive surface from a so-called {\tt hetTP} surrogate, a fitted heteroskedastic Student-$t$ process, is overlayed on the figure via mean and 95\% predictive intervals.  Although this fitted surface doesn't add much to the visualization here, its analog is essential in higher dimensions.

As an illustration, let's return now to $\nu \sim \mathrm{Gamma}(\alpha, \beta)$.  Our supplementary material provides a code which evaluates BFs on a space-filling maximin Latin hypercube sample \citep{morris1995exploratory} in $\alpha \times \beta$-space of size 40, using a recently updated version of the {\tt monomvn} library to accommodate the Gamma prior.  Five replicates are obtained at each input setting, for a total of 200 runs.  Performing these runs requires a similar computational expense to the earlier Exp experiment -- hundreds of hours rather than hundreds of days -- despite the higher dimension.
\begin{figure}[ht!]
\centering
\includegraphics[width=180mm,trim=0 0 0 0]{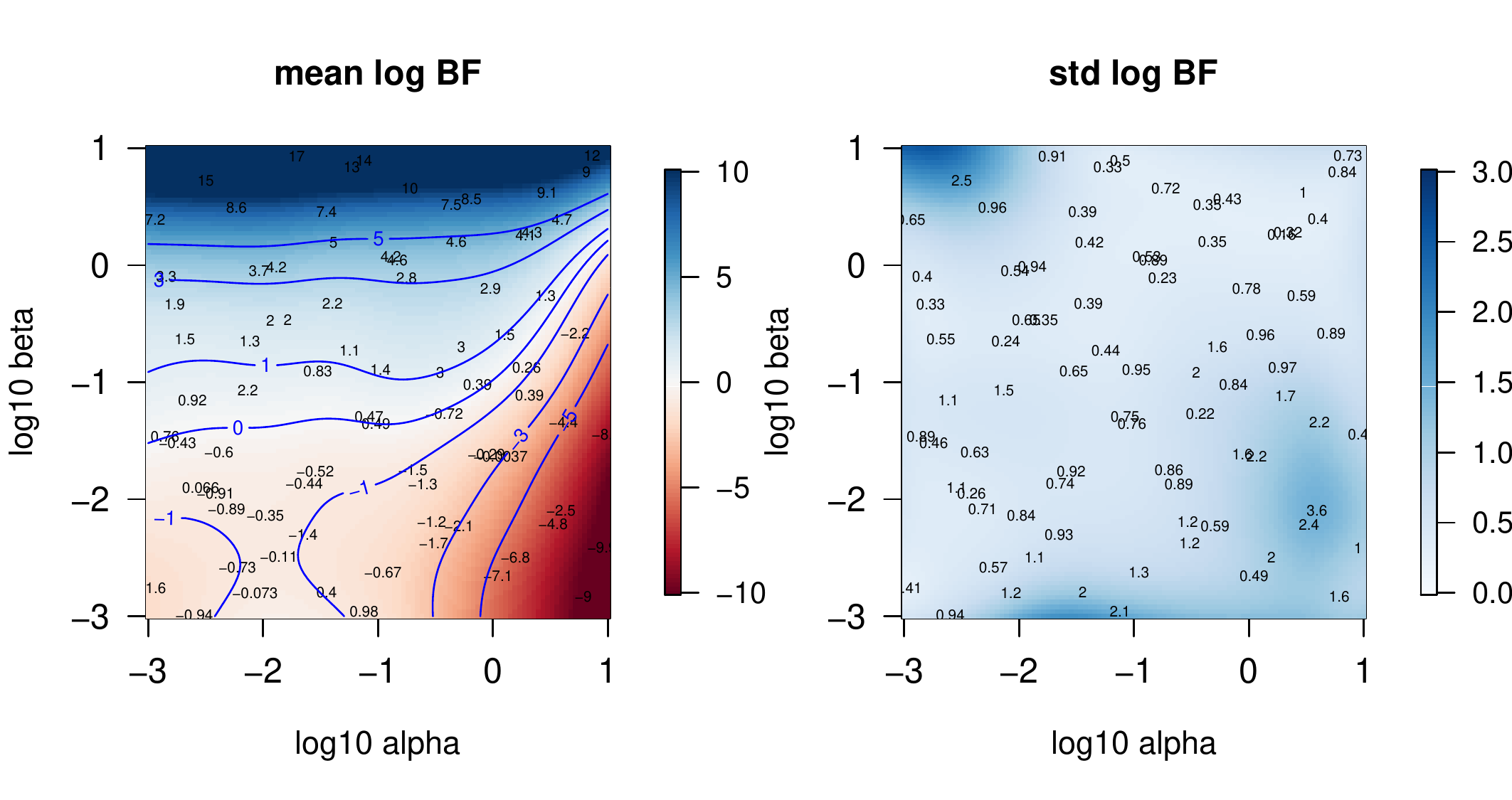}
\caption{Log Bayes factors for varying hyperparameter prior settings $\alpha$ and $\beta$ in $\nu \sim \mathrm{Gamma}(\alpha, \beta)$.  The $x$-axis ($\beta$) and $y$-axis are both shown in $\log_{10}$ space.  A predictive surface from {\tt hetTP} is overlayed via mean (left) and standard deviation (right).  The contours come from \cite{Kass1995}.}
\label{f:gammabfs}
\end{figure}
Figure \ref{f:gammabfs} shows the outcome of that experiment via a fitted {\tt hetTP} surface: mean on the left and standard deviation on the right.  The numbers overlayed on the figure are the average $\mathrm{BF}_{\mathcal{N}\mathrm{St}}$ obtained for the five replicates at each input location.  Observe that the story here is much the same as before in terms of $\beta$, which is mapped to $\theta$ in the earlier experiment, especially nearby $\alpha = 1$ (i.e., $\log_{10} \alpha = 0$) where the equivalence is exact.  The left panel shows that along this slice, one can realize just about any conclusion depending on hyperparameter specification.  Smaller $\alpha$ values tell a somewhat more nuanced story, however.  A rather large range of smaller $\alpha$ values lead to somewhat less sensitivity in the outcome due to the $\beta$ hyperparameter setting, except when $\beta$ is quite large.  It would appear that it is essential to have a small $\alpha$ setting if the data are going to have any influence on model selection via BF.  The right panel shows that the variance surface is indeed changing over the input space, justifying the heteroskedastic surrogate.

\section{Discussion}
\label{sec:discussion}

The instability of Bayes factors as a function of prior choices on parameters has been commented on in the statistical literature that develops Bayesian model comparison and selection. While the issue has been frequently mentioned, less attention has been spent on developing approaches to help researchers understand the extent of sensitivity in their own specific analyses. The two-fold purpose of this article has been to educate the broader statistical community about sensitivity in Bayes factors through interactive visualization, and to propose surrogate modeling in the form of computer surrogate models to examine the Bayes factor surface in cases where computing Bayes factors is too expensive to accomplish extensively.  We hope these contributions will provide researchers the insight and tools necessary to asses sensitivity in Bayes factors for their own specific research interests.

Much has been said and written about the downside of the near ubiquitous use of $p$-values for hypothesis testing across many branches of science. While it is unclear whether Bayesian model selection can or should ultimately supplant use of $p$-values on a broad scale, illustrating these sensitivities helps researchers use Bayes factors and Bayesian model selection responsibly.  Further, we hope availability of Bayes factor surface plots will help journal reviewers assess these sensitivities and  prevent statistical malpractice (intentional or otherwise) through exploitation of these sensitivities. Many of the concerns about $p$-values relate specifically to the practice of setting hard thresholds to declare results statistically significant or not \citep[see, e.g.,][]{Lakens2018,Benjamin2018}.  Similar thresholding of Bayesian metrics surely will not serve as a panacea to these concerns.

Whereas research papers used to include visuals of prior to posterior updating at the parameter level, and trace plots from MCMC chains to indicate good mixing, these are now pass\'{e}.  If they are not cut entirely, they are frequently relegated to an appendix.  The extra transparency such visuals provide are apparently no longer of value to the literature, perhaps because audiences have matured and readers/referees are now better able to sniff out troublesome modeling and inference choices through other, less direct, means.  Yet a similar setup is sorely needed for model selection.  Helpful visuals are not well-established and, as we show in our examples, it is easy to gain a false sense of security from the outcome of an experiment for which further analysis (e.g., visualization of a Bayes factor surface) reveals was actually balanced on a knife's edge.  Studies involving Bayesian model selection desperately need greater transparency.

We acknowledge that when many parameters are being tested simultaneously, it will be harder to visualize a single Bayes factor surface and multiple plots might be necessary. Individual surfaces can be helpful in facilitating lower-dimensional slices and projections.  However, the quality of that fitted surrogate will depend on the input design, with filling space becoming harder (as a function of a fixed design size) as the input dimension increases.  An attractive option here is sequential design.  \citet{Binois2018b} show how a minimum integrated mean-square prediction error criterion can be used to dynamically select design sites and degrees of replication in order to focus sampling on parts of the input space where the signal (e.g., in $\mathrm{BF}_{\mathcal{N}\mathrm{St}}$, say) is harder to extract from the noise.  Treating the Bayes factor surface surrogate as a sequential experiment in that context represents a promising avenue for further research. While this may seem potentially burdensome, we believe that the availability of Bayes factor surfaces will remain valuable.  We advise researchers to prioritize surfaces that address parameters that differ between the hypotheses.  Typically, hyperparameters for quantities common to both models affect Bayes factors to a much lesser degree than hyperparameters on the quantities being tested.  We encourage users to use the accompanying {\sf R} {\tt shiny} application to explore the comparatively large impact of changing $\phi$ and $\mu$ (both related to the parameter $\beta$ that is being tested) compared with the smaller influence $a$ and $b$ have, as these latter hyperparameters govern the error precision which is common to both models.

\bibliographystyle{Chicago}

\bibliography{refs}

\begin{thebibliography}{}

\bibitem[\protect\citeauthoryear{Andersen}{Andersen}{1957}]{andersen:1957}
Andersen, T. (1957, June).
\newblock Maximum likelihood estimates for a multivariate normal distribution
  when some observations are missing.
\newblock {\em J.~of the American Statistical Association\/}~{\em 52},
  200--203.

\bibitem[\protect\citeauthoryear{Bartlett}{Bartlett}{1957}]{Bartlett1957}
Bartlett, M.~S. (1957).
\newblock A comment on {D}. {V}. {L}indley's statistical paradox.
\newblock {\em Biometrika\/}~{\em 44\/}(3-4), 533--534.

\bibitem[\protect\citeauthoryear{Bates, M{\"a}chler, Bolker, and Walker}{Bates
  et~al.}{2015}]{Bates2015}
Bates, D., M.~M{\"a}chler, B.~Bolker, and S.~Walker (2015).
\newblock Fitting linear mixed-effects models using {lme4}.
\newblock {\em Journal of Statistical Software\/}~{\em 67\/}(1), 1--48.

\bibitem[\protect\citeauthoryear{Benjamin, Berger, Johannesson, Nosek,
  Wagenmakers, Berk, Bollen, Brembs, Brown, Camerer, Cesarini, Chambers, Clyde,
  Cook, De~Boeck, Dienes, Dreber, Easwaran, Efferson, Fehr, Fidler, Field,
  Forster, George, Gonzalez, Goodman, Green, Green, Greenwald, Hadfield,
  Hedges, Held, Hua~Ho, Hoijtink, Hruschka, Imai, Imbens, Ioannidis, Jeon,
  Jones, Kirchler, Laibson, List, Little, Lupia, Machery, Maxwell, McCarthy,
  Moore, Morgan, Munafó, Nakagawa, Nyhan, Parker, Pericchi, Perugini, Rouder,
  Rousseau, Savalei, Schönbrodt, Sellke, Sinclair, Tingley, Van~Zandt, Vazire,
  Watts, Winship, Wolpert, Xie, Young, Zinman, and Johnson}{Benjamin
  et~al.}{2018}]{Benjamin2018}
Benjamin, D.~J., J.~O. Berger, M.~Johannesson, B.~A. Nosek, E.-J. Wagenmakers,
  R.~Berk, K.~A. Bollen, B.~Brembs, L.~Brown, C.~Camerer, D.~Cesarini, C.~D.
  Chambers, M.~Clyde, T.~D. Cook, P.~De~Boeck, Z.~Dienes, A.~Dreber,
  K.~Easwaran, C.~Efferson, E.~Fehr, F.~Fidler, A.~P. Field, M.~Forster, E.~I.
  George, R.~Gonzalez, S.~Goodman, E.~Green, D.~P. Green, A.~G. Greenwald,
  J.~D. Hadfield, L.~V. Hedges, L.~Held, T.~Hua~Ho, H.~Hoijtink, D.~J.
  Hruschka, K.~Imai, G.~Imbens, J.~P.~A. Ioannidis, M.~Jeon, J.~H. Jones,
  M.~Kirchler, D.~Laibson, J.~List, R.~Little, A.~Lupia, E.~Machery, S.~E.
  Maxwell, M.~McCarthy, D.~A. Moore, S.~L. Morgan, M.~Munafó, S.~Nakagawa,
  B.~Nyhan, T.~H. Parker, L.~Pericchi, M.~Perugini, J.~Rouder, J.~Rousseau,
  V.~Savalei, F.~D. Schönbrodt, T.~Sellke, B.~Sinclair, D.~Tingley,
  T.~Van~Zandt, S.~Vazire, D.~J. Watts, C.~Winship, R.~L. Wolpert, Y.~Xie,
  C.~Young, J.~Zinman, and V.~E. Johnson (2018, January).
\newblock Redefine statistical significance.
\newblock {\em Nature Human Behaviour\/}~{\em 2\/}(1), 6--10.

\bibitem[\protect\citeauthoryear{Berger}{Berger}{1985}]{Berger1985}
Berger, J. (1985).
\newblock {\em {Statistical Decision Theory and Bayesian Analysis, second
  edition}}.
\newblock Springer.

\bibitem[\protect\citeauthoryear{Berger and Pericchi}{Berger and
  Pericchi}{1996}]{Berger1996}
Berger, J.~O. and L.~R. Pericchi (1996).
\newblock The intrinsic bayes factor for model selection and prediction.
\newblock {\em Journal of the American Statistical Association\/}~{\em
  91\/}(433), 109--122.

\bibitem[\protect\citeauthoryear{Binois and Gramacy}{Binois and
  Gramacy}{2018}]{hetGP}
Binois, M. and R.~B. Gramacy (2018).
\newblock {\em {\tt hetGP}: {H}eteroskedastic {G}aussian Process Modeling and
  Design under Replication}.
\newblock R package version 1.0.3.

\bibitem[\protect\citeauthoryear{Binois, Gramacy, and Ludkovski}{Binois
  et~al.}{2018}]{Binois2018}
Binois, M., R.~B. Gramacy, and M.~Ludkovski (2018).
\newblock Practical heteroscedastic {G}aussian process modeling for large
  simulation experiments.
\newblock {\em Journal of Computational and Graphical Statistics\/}~{\em
  27\/}(4), 808--821.

\bibitem[\protect\citeauthoryear{Binois, Huang, Gramacy, and Ludkovski}{Binois
  et~al.}{2019}]{Binois2018b}
Binois, M., J.~Huang, R.~B. Gramacy, and M.~Ludkovski (2019).
\newblock Replication or exploration? sequential design for stochastic
  simulation experiments.
\newblock {\em Technometrics\/}~{\em 61\/}(1), 7--23.

\bibitem[\protect\citeauthoryear{Boos and Stefanski}{Boos and
  Stefanski}{2011}]{Boos2011}
Boos, D.~D. and L.~A. Stefanski (2011).
\newblock P-value precision and reproducibility.
\newblock {\em The American Statistician\/}~{\em 65\/}(4), 213--221.

\bibitem[\protect\citeauthoryear{Chang, Cheng, Allaire, Xie, and
  McPherson}{Chang et~al.}{2017}]{Rshiny2017}
Chang, W., J.~Cheng, J.~Allaire, Y.~Xie, and J.~McPherson (2017).
\newblock {\em shiny: Web Application Framework for R}.
\newblock R package version 1.0.5.

\bibitem[\protect\citeauthoryear{Chipman, George, and McCulloch}{Chipman
  et~al.}{2001}]{Chipman2001}
Chipman, H., E.~I. George, and R.~E. McCulloch (2001).
\newblock {\em The Practical Implementation of Bayesian Model Selection},
  Volume Volume 38 of {\em Lecture Notes--Monograph Series}, pp.\  65--116.
\newblock Beachwood, OH: Institute of Mathematical Statistics.

\bibitem[\protect\citeauthoryear{Chung, Binois, Gramacy, Bardsley, Moquin,
  Smith, and Smith}{Chung et~al.}{2019}]{Chung2018}
Chung, M., M.~Binois, R.~B. Gramacy, J.~M. Bardsley, D.~J. Moquin, A.~P. Smith,
  and A.~M. Smith (2019).
\newblock Parameter and uncertainty estimation for dynamical systems using
  surrogate stochastic processes.
\newblock {\em SIAM Journal on Scientific Computing\/}~{\em 41\/}(4),
  A2212--A2238.

\bibitem[\protect\citeauthoryear{Fernandez, Ley, and Steel}{Fernandez
  et~al.}{2001}]{Fernandez2001}
Fernandez, C., E.~Ley, and M.~F. Steel (2001).
\newblock Benchmark priors for bayesian model averaging.
\newblock {\em Journal of Econometrics\/}~{\em 100\/}(2), 381--427.

\bibitem[\protect\citeauthoryear{Fonseca, Ferreira, and Migon}{Fonseca
  et~al.}{2008}]{Fonseca2008}
Fonseca, T. C.~O., M.~A.~R. Ferreira, and H.~S. Migon (2008).
\newblock Objective bayesian analysis for the student-t regression model.
\newblock {\em Biometrika\/}~{\em 95\/}(2), 325--333.

\bibitem[\protect\citeauthoryear{Foster, George, et~al.}{Foster
  et~al.}{1994}]{Foster1994}
Foster, D.~P., E.~I. George, et~al. (1994).
\newblock The risk inflation criterion for multiple regression.
\newblock {\em The Annals of Statistics\/}~{\em 22\/}(4), 1947--1975.

\bibitem[\protect\citeauthoryear{Geweke}{Geweke}{1992}]{geweke:1992}
Geweke, J. (1992).
\newblock Priors for microeconomic times series and their application.
\newblock Technical Report Institute of Empirical Macroeconomics Discussion
  Paper No.64, Federal Reserve Bank of Minneapolis.

\bibitem[\protect\citeauthoryear{Geweke}{Geweke}{1993}]{geweke:1993}
Geweke, J. (1993, Dec).
\newblock Bayesian treatment of the independent student--$t$ linear model.
\newblock {\em J.~of Applied Econometrics\/}~{\em Vol. 8, Supplement: Special
  Issue on Econometric Inference Using Simulation Techniques}, S19--S40.

\bibitem[\protect\citeauthoryear{Gramacy}{Gramacy}{2017}]{monomvn}
Gramacy, R.~B. (2017).
\newblock {\em {\tt monomvn}: Estimation for Multivariate Normal and
  {S}tudent-$t$ Data with Monotone Missingness}.
\newblock R package version 1.9-7.

\bibitem[\protect\citeauthoryear{Gramacy and Pantaleo}{Gramacy and
  Pantaleo}{2010}]{Gramacy2010}
Gramacy, R.~B. and E.~Pantaleo (2010, 06).
\newblock Shrinkage regression for multivariate inference with missing data,
  and an application to portfolio balancing.
\newblock {\em Bayesian Anal.\/}~{\em 5\/}(2), 237--262.

\bibitem[\protect\citeauthoryear{Hoff}{Hoff}{2009}]{Hoff2009}
Hoff, P. (2009).
\newblock {\em A First Course in Bayesian Statistical Methods}.
\newblock Springer Texts in Statistics. Springer New York.

\bibitem[\protect\citeauthoryear{Ioannidis}{Ioannidis}{2005}]{Ioannidis2005}
Ioannidis, J. P.~A. (2005, 08).
\newblock Why most published research findings are false.
\newblock {\em PLOS Medicine\/}~{\em 2\/}(8).

\bibitem[\protect\citeauthoryear{Jacquier, Polson, and Rossi}{Jacquier
  et~al.}{2004}]{jacq:polson:rossi:2004}
Jacquier, E., N.~Polson, and P.~E. Rossi (2004).
\newblock Bayesian analysis of stochastic volatility models with fat-tails and
  correlated errors.
\newblock {\em J.~of Econometrics\/}~{\em 122}, 185--212.

\bibitem[\protect\citeauthoryear{Jeffreys}{Jeffreys}{1935}]{Jeffreys1935}
Jeffreys, H. (1935).
\newblock Some tests of significance, treated by the theory of probability.
\newblock {\em Mathematical Proceedings of the Cambridge Philosophical
  Society\/}~{\em 31\/}(2), 203â€“222.

\bibitem[\protect\citeauthoryear{Jeffreys}{Jeffreys}{1961}]{Jeffreys1961}
Jeffreys, H. (1961).
\newblock {\em The Theory of Probability}, Volume~2 of {\em Oxford Classic
  Texts in the Physical Sciences}.
\newblock Oxford University Press.

\bibitem[\protect\citeauthoryear{Kass and Raftery}{Kass and
  Raftery}{1995}]{Kass1995}
Kass, R.~E. and A.~E. Raftery (1995).
\newblock Bayes factors.
\newblock {\em Journal of the American Statistical Association\/}~{\em
  90\/}(430), 773--795.

\bibitem[\protect\citeauthoryear{Kass and Wasserman}{Kass and
  Wasserman}{1995}]{KassWasserman1995}
Kass, R.~E. and L.~Wasserman (1995).
\newblock A reference bayesian test for nested hypotheses and its relationship
  to the schwarz criterion.
\newblock {\em Journal of the American Statistical Association\/}~{\em
  90\/}(431), 928--934.

\bibitem[\protect\citeauthoryear{Lakens, Adolfi, Albers, Anvari, Apps, Argamon,
  Baguley, Becker, Benning, Bradford, Buchanan, Caldwell, Van~Calster,
  Carlsson, Chen, Chung, Colling, Collins, Crook, Cross, Daniels, Danielsson,
  DeBruine, Dunleavy, Earp, Feist, Ferrell, Field, Fox, Friesen, Gomes,
  Gonzalez-Marquez, Grange, Grieve, Guggenberger, Grist, van Harmelen,
  Hasselman, Hochard, Hoffarth, Holmes, Ingre, Isager, Isotalus, Johansson,
  Juszczyk, Kenny, Khalil, Konat, Lao, Larsen, Lodder, Lukavský, Madan,
  Manheim, Martin, Martin, Mayo, McCarthy, McConway, McFarland, Nio, Nilsonne,
  de~Oliveira, de~Xivry, Parsons, Pfuhl, Quinn, Sakon, Saribay, Schneider,
  Selvaraju, Sjoerds, Smith, Smits, Spies, Sreekumar, Steltenpohl, Stenhouse,
  Świątkowski, Vadillo, Van~Assen, Williams, Williams, Williams, Yarkoni,
  Ziano, and Zwaan}{Lakens et~al.}{2018}]{Lakens2018}
Lakens, D., F.~G. Adolfi, C.~J. Albers, F.~Anvari, M.~A.~J. Apps, S.~E.
  Argamon, T.~Baguley, R.~B. Becker, S.~D. Benning, D.~E. Bradford, E.~M.
  Buchanan, A.~R. Caldwell, B.~Van~Calster, R.~Carlsson, S.-C. Chen, B.~Chung,
  L.~J. Colling, G.~S. Collins, Z.~Crook, E.~S. Cross, S.~Daniels,
  H.~Danielsson, L.~DeBruine, D.~J. Dunleavy, B.~D. Earp, M.~I. Feist, J.~D.
  Ferrell, J.~G. Field, N.~W. Fox, A.~Friesen, C.~Gomes, M.~Gonzalez-Marquez,
  J.~A. Grange, A.~P. Grieve, R.~Guggenberger, J.~Grist, A.-L. van Harmelen,
  F.~Hasselman, K.~D. Hochard, M.~R. Hoffarth, N.~P. Holmes, M.~Ingre, P.~M.
  Isager, H.~K. Isotalus, C.~Johansson, K.~Juszczyk, D.~A. Kenny, A.~A. Khalil,
  B.~Konat, J.~Lao, E.~G. Larsen, G.~M.~A. Lodder, J.~Lukavský, C.~R. Madan,
  D.~Manheim, S.~R. Martin, A.~E. Martin, D.~G. Mayo, R.~J. McCarthy,
  K.~McConway, C.~McFarland, A.~Q.~X. Nio, G.~Nilsonne, C.~L. de~Oliveira,
  J.-J.~O. de~Xivry, S.~Parsons, G.~Pfuhl, K.~A. Quinn, J.~J. Sakon, S.~A.
  Saribay, I.~K. Schneider, M.~Selvaraju, Z.~Sjoerds, S.~G. Smith, T.~Smits,
  J.~R. Spies, V.~Sreekumar, C.~N. Steltenpohl, N.~Stenhouse, W.~Świątkowski,
  M.~A. Vadillo, M.~A. L.~M. Van~Assen, M.~N. Williams, S.~E. Williams, D.~R.
  Williams, T.~Yarkoni, I.~Ziano, and R.~A. Zwaan (2018, March).
\newblock Justify your alpha.
\newblock {\em Nature Human Behaviour\/}~{\em 2\/}(3), 168--171.

\bibitem[\protect\citeauthoryear{Lavine and Schervish}{Lavine and
  Schervish}{1999}]{Lavine1999}
Lavine, M. and M.~J. Schervish (1999).
\newblock Bayes factors: What they are and what they are not.
\newblock {\em The American Statistician\/}~{\em 53\/}(2), 119--122.

\bibitem[\protect\citeauthoryear{Liang, Paulo, Molina, Clyde, and Berger}{Liang
  et~al.}{2008}]{Liang2008}
Liang, F., R.~Paulo, G.~Molina, M.~A. Clyde, and J.~O. Berger (2008).
\newblock Mixtures of g priors for {B}ayesian variable selection.
\newblock {\em Journal of the American Statistical Association\/}~{\em
  103\/}(481), 410--423.

\bibitem[\protect\citeauthoryear{Lindley}{Lindley}{1957}]{Lindley1957}
Lindley, D.~V. (1957).
\newblock A statistical paradox.
\newblock {\em Biometrika\/}~{\em 44\/}(1/2), 187--192.

\bibitem[\protect\citeauthoryear{Markowitz}{Markowitz}{1959}]{markowitz:1959}
Markowitz, H. (1959).
\newblock {\em Portfolio Selection: Efficient Diversification of Investments}.
\newblock New York: John Wiley.

\bibitem[\protect\citeauthoryear{McKay, Conover, and Beckman}{McKay
  et~al.}{1979}]{mcka:cono:beck:1979}
McKay, M.~D., W.~J. Conover, and R.~J. Beckman (1979).
\newblock A comparison of three methods for selecting values of input variables
  in the analysis of output from a computer code.
\newblock {\em Technometrics\/}~{\em 21}, 239--245.

\bibitem[\protect\citeauthoryear{Morris and Mitchell}{Morris and
  Mitchell}{1995}]{morris1995exploratory}
Morris, M.~D. and T.~J. Mitchell (1995).
\newblock Exploratory designs for computational experiments.
\newblock {\em Journal of statistical planning and inference\/}~{\em 43\/}(3),
  381--402.

\bibitem[\protect\citeauthoryear{O'Hagan}{O'Hagan}{1995}]{Ohagan1995}
O'Hagan, A. (1995).
\newblock Fractional {B}ayes factors for model comparison.
\newblock {\em Journal of the Royal Statistical Society. Series B
  (Methodological)\/}~{\em 57\/}(1), 99--138.

\bibitem[\protect\citeauthoryear{Ormerod, Stewart, Yu, and Romanes}{Ormerod
  et~al.}{2017}]{Ormerod2017}
Ormerod, J.~T., M.~Stewart, W.~Yu, and S.~E. Romanes (2017).
\newblock Bayesian hypothesis tests with diffuse priors: Can we have our cake
  and eat it too?
\newblock {\em arXiv preprint arXiv:1710.09146\/}.

\bibitem[\protect\citeauthoryear{Santner, Williams, and Notz}{Santner
  et~al.}{2018}]{sant:will:notz:2003}
Santner, T.~J., B.~J. Williams, and W.~I. Notz (2018).
\newblock {\em The Design and Analysis of Computer Experiments\/} (2 ed.).
\newblock New York, NY: Springer-Verlag.

\bibitem[\protect\citeauthoryear{Schwarz}{Schwarz}{1978}]{Schwarz1978}
Schwarz, G. (1978, 03).
\newblock Estimating the dimension of a model.
\newblock {\em Ann. Statist.\/}~{\em 6\/}(2), 461--464.

\bibitem[\protect\citeauthoryear{Stambaugh}{Stambaugh}{1997}]{stambaugh:1997}
Stambaugh, R.~F. (1997).
\newblock Analyzing investments whose histories differ in lengh.
\newblock {\em J.~of Financial Economics\/}~{\em 45}, 285--331.

\bibitem[\protect\citeauthoryear{Trafimow and Marks}{Trafimow and
  Marks}{2015}]{Trafimow2015}
Trafimow, D. and M.~Marks (2015).
\newblock Editorial.
\newblock {\em Basic and Applied Social Psychology\/}~{\em 37\/}(1), 1--2.

\bibitem[\protect\citeauthoryear{Wasserstein, Lazar, et~al.}{Wasserstein
  et~al.}{2016}]{Wasserstein2016}
Wasserstein, R.~L., N.~A. Lazar, et~al. (2016).
\newblock The {ASA}’s statement on p-values: context, process, and purpose.
\newblock {\em The American Statistician\/}~{\em 70\/}(2), 129--133.

\bibitem[\protect\citeauthoryear{Yao, Vehtari, Simpson, Gelman, et~al.}{Yao
  et~al.}{2018}]{Yao2018}
Yao, Y., A.~Vehtari, D.~Simpson, A.~Gelman, et~al. (2018).
\newblock Using stacking to average {B}ayesian predictive distributions (with
  discussion).
\newblock {\em Bayesian Analysis\/}~{\em 13\/}(3), 917--1003.

\bibitem[\protect\citeauthoryear{Young and Karr}{Young and
  Karr}{2011}]{Young2011}
Young, S.~S. and A.~Karr (2011).
\newblock Deming, data and observational studies.
\newblock {\em Significance\/}~{\em 8\/}(3), 116--120.

\bibitem[\protect\citeauthoryear{Zellner}{Zellner}{1986}]{Zellner1986}
Zellner, A. (1986).
\newblock On assessing prior distributions and {B}ayesian regression analysis
  with g-prior distributions.
\newblock In {\em Bayesian Inference and Decision Techniques: Essays in Honor
  of Bruno de Finetti}, Chapter~15, pp.\  233--243. New York: Elsevier Science
  Publishers, Inc.

\bibitem[\protect\citeauthoryear{Zellner and Siow}{Zellner and
  Siow}{1980}]{Zellner1980}
Zellner, A. and A.~Siow (1980, Feb).
\newblock Posterior odds ratios for selected regression hypotheses.
\newblock {\em Trabajos de Estadistica Y de Investigacion Operativa\/}~{\em
  31\/}(1), 585--603.

\end{thebibliography}

\appendix

\section*{Appendix A: Further regression details}
\label{app:regress}

The Bayesian model for the mixture $g$-prior comes from \cite{Liang2008}:
\begin{align*}
y_i|\alpha, \beta, \gamma, M_k, &\stackrel{\mathrm{iid}}{\sim} \mathcal{N} \left(\alpha+\beta x_i, \frac{1}{\gamma} \right) \\
\beta|\gamma, g  &\sim \mathcal{N} \left(0,\frac{g}{\gamma\sum_{i=1}^nx_i^2}\right) \\
p \big(\alpha,\gamma \big) &= \frac{1}{\phi}  \\
g &\sim \mathrm{IG} \left( \frac{1}{2},\frac{n}{2} \right) 
\end{align*}
The marginal likelihood can be obtained by analytically integrating $\alpha$, $\beta$, and $\gamma$, then approximating the integral over $g$.  As in the original paper, we use Laplace approximation for this last integral.

Fractional Bayes factors: Let $m$ denote a training sample size, $n$ the sample size, and $b=\frac{m}{n}$. The likelihood raised to $b$ approximates the likelihood of a training sample, a fact used to form fractional Bayes factors. Fractional Bayes factors are appealing since there is no need to choose an arbitrary training sample or spend computational resources averaging over some or all possible training samples to obtain an intrinsic Bayes factor \citep{Berger1996}. The prior distribution for the fractional Bayes factor analysis is the reference prior: 
$$
p(\alpha,\beta,\gamma) = \frac{c}{\gamma} 
$$
where $c$ is a constant.  This prior is improper, though it yields a proper posterior distribution for inference on parameters and  is eligible for model selection due to the availability of fractional Bayes factors. The minimal training sample in this study is $m=3$.  Further details about fractional Bayes factors are left to the literature.

\section*{Appendix B: Further hierarchical linear model details}
\label{app:hlm}

The ``slopes and intercepts'' model $M_1$ has school-specific slopes and intercepts, thus $\beta_j$ has a length of 2 (i.e. $p_1=2$), thus $X_j$ is a school-specific model matrix for simple linear regression.  For the "means only" model $M_2$, $X_j$ is an $n_j \times 1$ column vector of ones and $p_2=1$.

In this example we have imposed a fixed $g$-prior on the school-level regression effects to make the integration for the marginal likelihood analytical except for one dimension.   Let $j=1,\ldots,m$, $n_j$ is the number of students in school $j$, $y_j$ is an $n_j$ dimensional vector of outcomes for school $j$, $X_j$ is the $n_j \times p$ model matrix for the $j^\mathrm{th}$ school, and $\beta_j$ is the vector of linear effects for the $j^\mathrm{th}$ school. Let $N=\sum_{j=1}^{m} n_j$. The marginal likelihood for this example is:
\begin{align*}
P(\bm{y}&|M_k)= (2\pi)^{-\frac{1}{2}N}\text{det}(\Lambda_0)^{-\frac{1}{2}} (g+1)^{-\frac{mp_k}{2}} \frac{(\frac{\nu_0 \sigma_0^2}{2})^{\nu_0/2}}{\Gamma(\frac{\nu_0}{2})} \exp\Big\{-\frac{1}{2}\mu_0^\top\Lambda_0^{-1}\mu_0\Big\} \times \\
 & \int\gamma^{\frac{1}{2}N + \frac{\nu_0}{2}-1} \text{det}\Big(\gamma S_2 + \Lambda_0^{-1} \Big)^{-1} \times \\
  &  \exp\Big\{ -\frac{\gamma}{2}s_3 - \frac{\nu_0\sigma^2_0}{2}\gamma + \frac{1}{2}\big(-\frac{\gamma}{2}S_1 + \mu_0^\top \Lambda_0^{-1} \big)\big(\gamma S_2 + \Lambda_0^{-1} \big)^{-1} \big(-\frac{\gamma}{2}S_1^\top + \Lambda_0^{-1}\mu_0 \big)\Big\}d\gamma
\end{align*}
where $S_1=\frac{-2}{g+1}\sum_{j=1}^m y_j^\top X_j$, $S_2=\frac{1}{g+1}\sum_{j=1}^m X_j^\top X_j$, and 
$s_3=\sum_{j=1}^m\Big[ y_j^\top y_j - \frac{g}{g+1}\big(y_j^\top X_j \hat{\beta}_j   \big)  \big]$. The integration with respect to $\gamma$ was performed using the trapezoid rule on the log precision scale.

Since the math scores are centered within each school, the y-intercept stored in the first element of $\beta_j$ is the school-level average.  This is the justification for having a common set of hyperparameters for the school level means in the ``means only'' model and school level y-intercepts in the ``slopes and intercepts'' model.

The term $(g+1)^{-\frac{mp_k}{2}}$ in $P(\bm{y}|M_k)$ mostly explains the extraordinary role $g$ plays influencing the Bayes factor in this setting. In these hierarchical models, $p_1=200$ and $p_2=100$. In a typical subset selection problem for regression, the difference in the number of parameters would usually be pretty small compared to $p_1$ and $p_2$ in these hierarchical linear models.  When the Bayes factor is formed, the resulting term is $(g+1)^{-5000}$ for these data. A graphical analysis (not shown) confirms that this term plays a much larger role than the other factors in $P(\bm{y}|M_k)$, and explains why the Bayes factor for this model depends so heavily on $g$.

\end{document}